\documentstyle[aps,prl]{revtex}
\begin{document}
\draft \twocolumn[\hsize\textwidth\columnwidth\hsize\csname
@twocolumnfalse\endcsname 
\title{New quantum phase transitions in the two-dimensional
$J_{1}-J_{2}$ model }
\author{O. P. Sushkov, J. Oitmaa,  and
 Zheng Weihong} \address{School of Physics, University of New South
 Wales, Sydney 2052, Australia}

\maketitle
\begin{abstract}
We analyze the phase diagram of the frustrated Heisenberg antiferromagnet,
the $J_{1}-J_{2}$ model, in two dimensions. 
Two quantum phase transitions in the model are already known: 
the second order transition from the N\'eel state to the spin liquid state
at $(J_2/J_1)_{c2}=0.38$, and the first order transition from the spin liquid state
to the collinear state at $(J_2/J_1)_{c4}=0.60$.
We have found evidence for two new second order phase transitions: the transition from the
spin columnar dimerized state to the state with plaquette type modulation at 
$(J_2/J_1)_{c3}=0.50\pm 0.02$,
and the transition from the simple N\'eel state to the N\'eel state with spin columnar
dimerization at  $(J_2/J_1)_{c1}=0.34\pm 0.04$.
We also present an independent calculation of $(J_2/J_1)_{c2}=0.38$ using a new approach. 
\end{abstract}

\pacs{PACS: 75.10.Jm, 75.30.Kz, 75.40.Gb, 75.30.Ds} ]

The nature of the quantum disordered phases of
 low-dimensional quantum antiferromagnets is a topic of fundamental
 importance for the physics of quantum magnetism \cite{subir}.  Such
 phases can result from mobile holes in an antiferromagnetic
 background as in the $t-J$ or Hubbard model at finite doping.
 Alternatively, competition of purely magnetic interactions can also
 lead to destruction of long-range order.  A typical example of the
 second kind is the $J_{1}-J_{2}$ model which exhibits a quantum
 disordered (spin-liquid) phase due to second-neighbor frustrating
 interactions.
Even though it has been intensively studied during the
 last ten years, the $J_{1}-J_{2}$ model apparently still holds many secrets.
This model is also an ideal testing ground for the theory of quantum
phase transitions because it has very complex dynamics and
contains a variety of transitions.  Exact
 diagonalization studies \cite{exact} have shown that the excitation
 spectrum of the model is quite complex and that finite-size effects
 are large \cite{exact1}. Spin-wave like expansions around the simple N\'eel
 state (which occurs for small frustration) naturally cannot give any
 information about the ground state at stronger frustration, and
 consequently non-perturbative methods are needed to analyze the
 latter regime.

  An important insight into the disordered regime was achieved by
 field-theory methods \cite{Read,us} and dimer series expansions
 \cite{series,us,series2}.
  The above works have established the range of the disordered regime,
  $0.38 < g < 0.60$ ($g = J_2/J_1$), and have also shown that the ground
 state in this regime is dominated by short-range singlet
 (dimer) formation in a given pattern (see Fig.1). The stability of
 such a configuration implies that the lattice symmetry is
 spontaneously broken and the ground state is four-fold
 degenerate. This picture is somewhat similar to the
one dimensional situation, where the Lieb-Schultz-Mattis
 theorem guarantees that a gapped phase always breaks the
 translational symmetry and is doubly degenerate, whereas gapless
 excitations correspond to a unique ground state \cite{LSM}.

 Two very recent calculations \cite{Capriotty,Jongh} performed by
 Green function Monte Carlo methods have raised new questions on the 
structure of the intermediate phase.
The authors of Ref. \cite{Capriotty} claim stability of the ``plaquette RVB''
state at $g \approx 0.5$. Reference \onlinecite{Jongh} comes to a different 
conclusion: there is a columnar spin dimerized state with plaquette type
modulation along the columns. An additional very interesting observation \cite{Jongh}
is that the columnar spin dimerization penetrates into the N\'eel phase to
$g \approx 0.3$. To conclude the list of observations which do not agree with
a simple spin liquid with columnar dimerization we  mention the divergence 
in the plaquette susceptibility found in Ref. \cite{series2} at $g \sim 0.5$.

In the present paper we elucidate all the above questions and come to the 
conclusion that two additional  quantum critical points exist in the phase
diagram of the system. These  critical points correspond to a new generic type
of  second order quantum phase transition considered in Ref. \onlinecite{Mike}. 
At each of the critical points points there is condensation of some singlet 
excitation and the critical dynamics is described by the nonlinear O(1) $\sigma$-model.

The Hamiltonian of the $J_{1}-J_{2}$ model reads:

\begin{equation}
\label{ham}
H = J_{1} \sum_{nn}{\bf S}_{i}\cdot {\bf S}_{j} + J_{2} \sum_{nnn} {\bf
  S}_{i}\cdot {\bf S}_{j},
\end{equation}
where $J_{1}$ is the nearest-neighbor, and $J_{2}$ is the frustrating
  next-nearest-neighbor Heisenberg exchange on a square lattice (see
  Fig.1).  Both couplings are antiferromagnetic, i.e.  $J_{1,2}>0$ and
  the spins $S_{i}=1/2$.  We also use the notation $g =J_2/J_1$.
The spin columnar dimerization at $g > g_{c2}$ is
well established \cite{Read,us,series,series2} and therefore we start
our consideration from this state shown schematically in Fig.1.
If there is an instability with respect to some kind of additional ordering
then the gap in the spectrum of some singlet excitation must vanish at the
corresponding critical point \cite{Mike}. We do not have a reliable technique 
for direct calculation of the singlet gap, but we do have a well developed series 
expansion technique for  calculation of static susceptibilities.
A static susceptibility is proportional to the corresponding Green's function at
zero frequency
\begin{equation}
\label{chi}
\chi_{\bf q} \propto G_{\bf q}(\omega=0) \sim Z_{\bf q}/\omega_{\bf q}^2,
\end{equation}
where $\omega_{\bf q}$ is the quasiparticle energy, and $Z_{\bf q}$ is the
quasiparticle residue.
So at the critical point  $1/\chi$ must vanish approximately as 
$(g-g_c)^{\gamma}$, with $\gamma \simeq 2(\nu-\eta)$, where $\nu$ is the critical 
index for  the spectral gap, $\Delta \propto (g-g_c)^{\nu}$, and $2\eta$ is
the critical index for the quasiparticle residue, $Z \propto (g-g_c)^{2\eta}$

To analyze possible plaquette type modulation  we calculate the susceptibility of
the spin columnar dimerized state with respect to the field \cite{series2}
\begin{equation}
\label{FP}
F_P=\sum_{i,j}(-1)^{j}{\bf S}_{i,j}\cdot {\bf S}_{i,j+1},
\end{equation}
which breaks the translational symmetry in the direction perpendicular to the
dimers. The series has been computed up to the seventh order in the dimerization 
parameter. Results for $1/\chi_P$ are shown in Fig. 2. The value of
 $1/\chi_P$ vanishes at $g_{c3}=0.50\pm 0.02$ and this is the critical
point for the second order quantum phase transition from a simple columnar
dimerized state to the 8-fold degenerate
columnar dimerized state with plaquette type bond modulation
in the direction perpendicular to the dimers suggested in Ref. \onlinecite{Jongh}.
This phase transition is of the generic type considered in Ref. \cite{Mike}
and therefore it is described by 2D nonlinear O(1) $\sigma$-model. 
The critical indexes for this model are \cite{NL}: $\nu \approx 0.630$, 
$\eta \approx 0.034$.
Therefore one shall expect $\gamma = 2(\nu-\eta) \approx 1.19$.
On the other hand the Dlog Pad\'e approximants to the series $\chi_P$ 
give $\gamma = 0.9 \pm 0.1$.
This is fair agreement, and we offer an explanation for the small discrepancy.
The phase transition is related to the condensation of some singlet excitation
which can be considered as a bound state of triplet excitations.
\begin{equation}
|s\rangle =a_2|tt\rangle + a_3|ttt\rangle + a_4|tttt\rangle + ...
\end{equation}
We would like to stress that there is very strong mixing between two-triplet and 
multi-triplet bound states. This mixing was the reason why vanishing of the singlet gap 
at $g=g_{c3}$ was missed in Reference  \cite{us}. In Ref.  \cite{us} 
analysis of the singlet excitation was based on a two-particle Bethe-Salpeter equation 
with further  account of multi-particle contributions as a small perturbation.
This assumption was wrong because of the strong mixing. So  at $g=g_{c3}$ we 
have condensation of effectively a multi-particle bound state with relatively 
small two particle component. The mixing between two-particle and multi-particle
components of the singlet excitation varies with $g$ and this effect
cannot be taken into account in the non-linear $\sigma$-model which assumes
condensation of an ``elementary'' (=structureless) field.
Ultimately very close to the critical point the variation of the
mixing can be neglected and one shall expect restoration of the pure $\sigma$-model
behavior. However it happens in so narrow vicinity of the critical point that the
present numerical data cannot assess it.

Let us consider now the appearance of spin dimer order at $g =g_{c1}$
as $g$ is increased from small values. 
A scenario put forward some 
time ago \cite{Read} and based on the analysis of the $Sp(N)$, $N \to \infty$ field 
theory suggests that the dimer order appears simultaneously with disappearance of the 
N\'eel order, $g_{c1}=g_{c2}$. The dynamics in the vicinity of the critical
point is described by the nonlinear O(3) $\sigma$-model in spite of an additional
dimer order parameter. The additional gapless excitation is irrelevant to the critical
dynamics because  this excitation has extremely large size \cite{Read}: 
$r \sim 1/(g-g_c)^M$, $M \gg 1$.
Another possibility is that $g_{c1} < g_{c2}$ and hence there are two
separated quantum phase transitions \cite{com1}. 
The transition at $g_{c2}=0.38$ is still
described by the nonlinear O(3) $\sigma$-model, while the transition at  $g_{c1}$ is
of the O(1)$\times$O(1)-type. So in the vicinity of the point $g_{c1}$ there is an
additional effective singlet field which can condense either at momentum
${\bf k}=(\pi,0)$ or $(0,\pi)$. The sign degeneracy of the scalar field
together with the momentum degeneracy gives a four-fold degenerate ground 
state which exactly corresponds to the  degeneracy of the spin-dimerized 
state.
A recent work based on the Green function Monte Carlo method \cite{Jongh}
gives a hint in favor of this picture.

Let us give the precise meaning to the terms relevant and irrelevant singlet excitation. 
We consider a quantum critical point at which the singlet gap $\Delta_s$ vanishes. 
An external field which is coupled to the singlet excitation, $\langle s|F|0\rangle \ne 0$,
is applied to the system.
If the corresponding susceptibility given by eq. (\ref{chi}) is diverging at the critical
point we call this singlet excitation ``relevant''. If the susceptibility is not
diverging we call the singlet excitation ``irrelevant''.
It is clear that for an irrelevant excitation the quasiparticle residue $Z$ vanishes
faster than $\Delta_s^2$.

To analyze the problem of spin dimer order we calculate the susceptibility of the N\'eel state 
with respect to the external  field which probes spin columnar dimerization.
\begin{equation}
\label{FD}
F_D=\sum_{i,j}(-1)^{i} {\bf S}_{i,j}\cdot {\bf S}_{i+1,j}.
\end{equation}
In this calculation we use the usual Ising series expansion up to seventh
order. Note that in spite of the similarity between (\ref{FP}) and (\ref{FD})
these are two quite different situations. The field (\ref{FP}) assumes that
the dimers aligned along the i-direction already exist and it probes a possible
modulation in the j-direction. The field (\ref{FD}) is applied to the N\'eel state and
therefore it does not assume any dimer order. The values of $1/\chi_D$ versus
$g$ are plotted in Fig. 3. It is clear that $1/\chi_D$ vanishes somewhere in the
interval $g \in [0.3,0.4]$, but  the data is not precise enough to distinguish
$g_{c1}$ from $g_{c2}$.
To distinguish between the two scenarios discussed above we have to realize
that Fig. 3 clearly indicates the {\it relevant}  singlet excitation.
In the case of an {\it irrelevant} singlet \cite{Read} the quasiparticle residue 
is extremely small, $Z\propto (g-g_c)^M$, $M \gg 1$
and hence the susceptibility has no divergence at the critical point.
Thus we conclude from Fig. 3 that $g_{c1}=0.34 \pm 0.04$, and that
$g_{c1} < g_{c2}$, so there is a region $g_{c1} < g < g_{c2}$
where the spin columnar dimer order and the N\'eel order coexist.
The critical dynamics at $g_{c1}$ is described by the {\it relevant} gapless
singlet excitation. There is no doubt that the {\it irrelevant}
gapless singlet excitation at $g \approx g_{c2}$ also exists, but it
has an exponentially small residue \cite{Kot1} and hence its contribution to
the susceptibility is negligible.

The final result we report here is a new way of estimating
$g_{c2} \simeq 0.38$. The previous best calculation \cite{us} 
was based on vanishing of the triplet gap in the spin liquid
phase. A previous attempt\cite{M} to estimate $g_{c2} $ by
 Ising expansions for the
staggered magnetization in the N\'eel phase  showed the magnetization
vanishing around 0.4, but the series were erratic in this region and the
precision low. The new estimate is based on Ising expansions\cite{M}
in the N\'eel phase for the 1st, 2nd and 3rd neighbor correlators 
$\langle S_i^x S_j^x\rangle $ and $\langle S_i^z S_j^z\rangle $,
where $z$ is the direction of staggered magnetization.
The series has been computed up to order 9 for 1st
and  2nd  neighbor correlators and to order 7 for 3rd neighbor correlator.
The differences of these correlators are shown in Fig.4. The transition point is
identified by the condition
$\langle S_i^x S_j^x\rangle = \langle S_i^z S_j^z\rangle $,
corresponding to restoration of spin rotational symmetry to the
ground state. This gives $g_{c2} \simeq 0.38(3)$,
in excellent agreement with previous results.

In conclusion, the  zero temperature phase diagram and the excitation spectra of the 
$J_1-J_2$ model are shown schematically in Fig. 5. There are four critical points:
$g_{c1}=0.34\pm 0.04$, $g_{c2}=0.38$, $g_{c3}=0.50 \pm 0.02$,
$g_{c4}=0.60$. The states are: $g < g_{c1}$ - the simple N\'eel state,
$g_{c1} < g < g_{c2}$ - the columnar dimerized  N\'eel state,
$g_{c2} < g < g_{c3}$ - the simple columnar dimerized spin liquid,
$g_{c3} < g < g_{c4}$ - the columnar dimerized spin liquid with
plaquette type modulation, $g > g_{c4}$ - the collinear state.
The transitions at $g_{c1}$ and $g_{c3}$ are second order phase transitions
of the O(1)$\times$O(1) and O(1) symmetry classes correspondingly.
Energies of the {\it relevant} singlet excitations vanish at the critical points.
The transition at $g_{c2}$ is a second order phase transition
described by the nonlinear O(3) $\sigma$-model. The energy of the triplet excitation
vanishes at $g \le g_{c2}$; at the critical point there is also a singlet excitation 
with zero gap, but this singlet is {\it irrelevant}.
The transition at $g_{c4}$ is probably of first order, but is very
close to  second order.

This work forms part of a research project supported by a grant from the 
Australian Research Council. 
We thank V. N. Kotov and J. M. J. van Leeuwen for very helpful stimulating
discussions. We are also grateful to S. Sachdev for very important comments.
The computation has been performed on 
Silicon Graphics Power Challenge and Convex machines.  We thank the New 
South Wales Centre for Parallel Computing for facilities and assistance 
with the calculations.

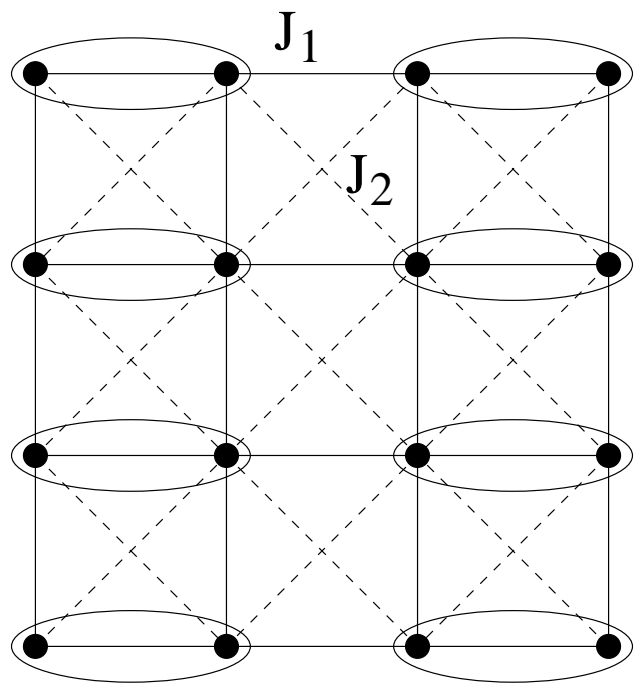
\begin{figure}
\caption
{Schematic picture of the simple columnar dimerized state.
The ovals represent two spins coupled into a singlet.}
\label{fig.1}
\end{figure}

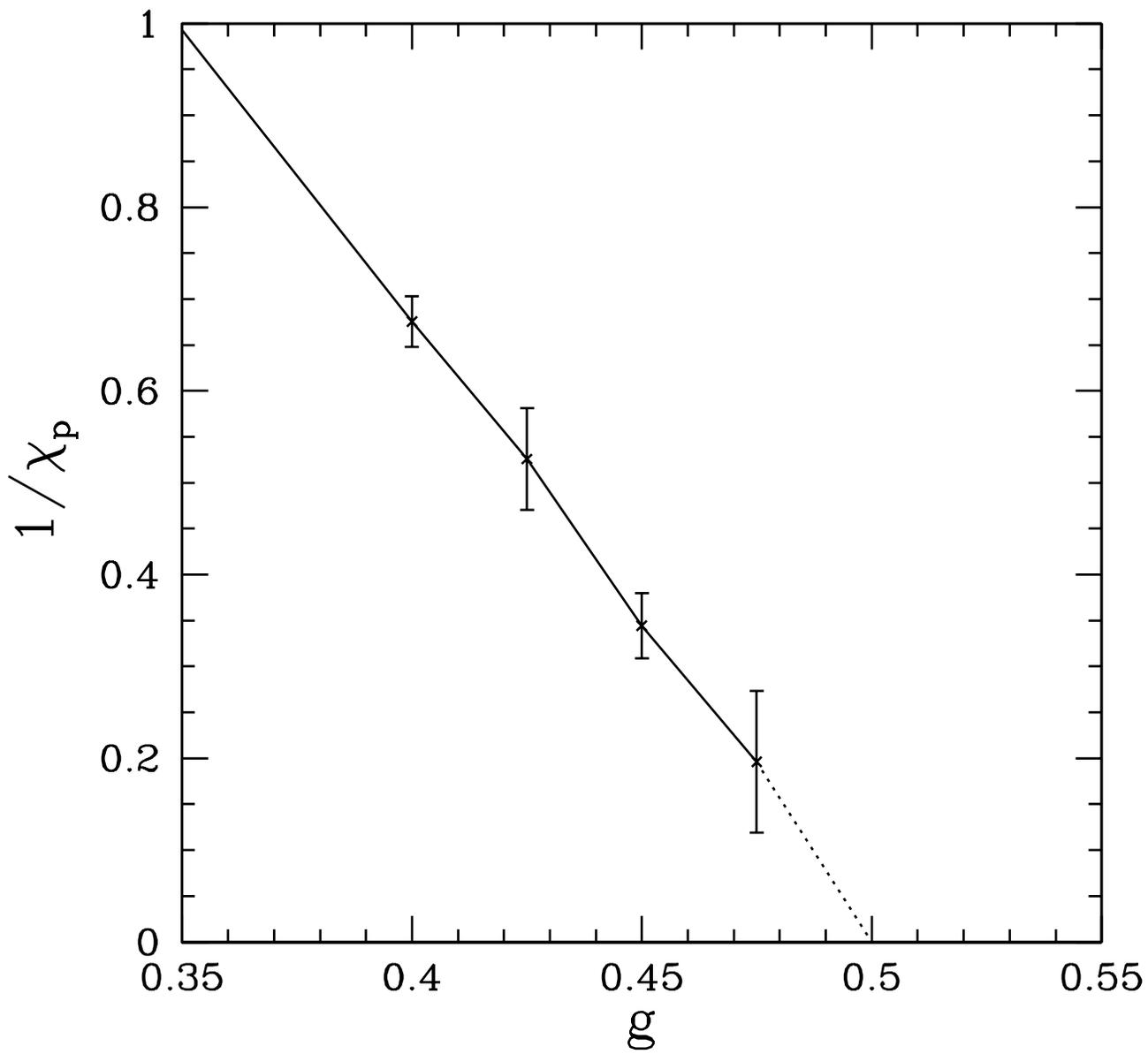
\begin{figure}
\caption
{The plot of $1/\chi_P$, where $\chi_P$ is the plaquette susceptibility
calculated in the simple columnar dimerized state using dimer series expansion. 
The value of $1/\chi_P$ vanishes at $g_{c3}=0.50\pm0.02$ indicating a transition 
to the columnar dimerized state with plaquette type modulation.}
\label{fig.2}
\end{figure}

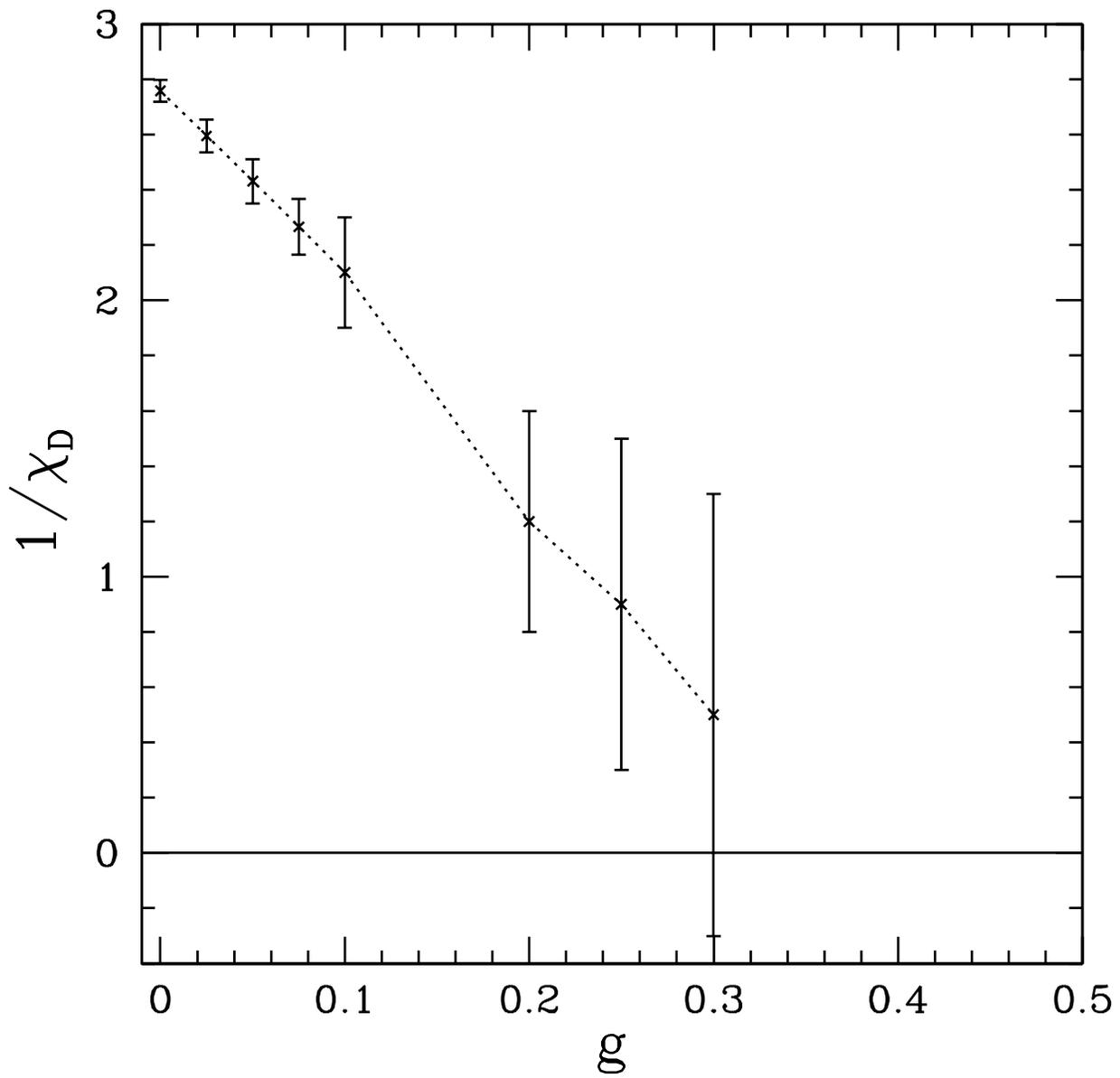
\begin{figure}
\caption
{The plot of $1/\chi_D$, where $\chi_D$ is the dimer susceptibility
calculated in the simple N\'eel state using Ising series expansion. The value of 
$1/\chi_D$ vanishes at $g_{c1}=0.34\pm 0.04$ indicating transition to the N\'eel
state with spontaneous spin columnar dimerization.}
\label{fig.3}
\end{figure}

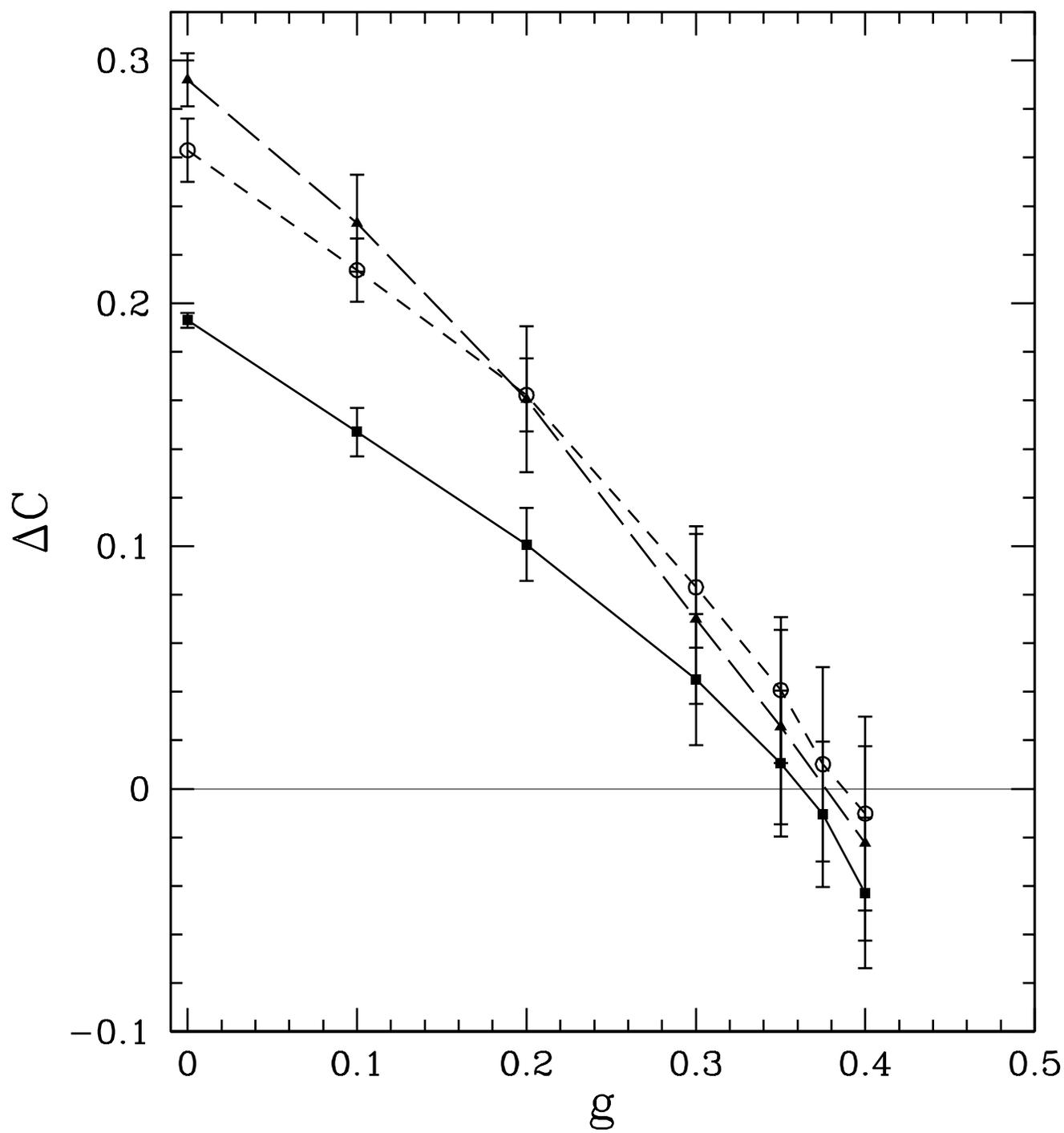
\begin{figure}
\caption
{Plot of the difference $\Delta C = 2 \vert \langle {\bf S}_i \cdot {\bf S}_j \rangle -
3 \langle S_i^z S_j^z \rangle \vert$ for 1st neighbors (full line), 2nd neighbors
(short dashed line), and 3rd neighbors (long-dashed line) versus $g = J_2/J_1$.
The full rotation symmetry of the ground state is restored when $\Delta C =0$.}
\label{fig.4}
\end{figure}

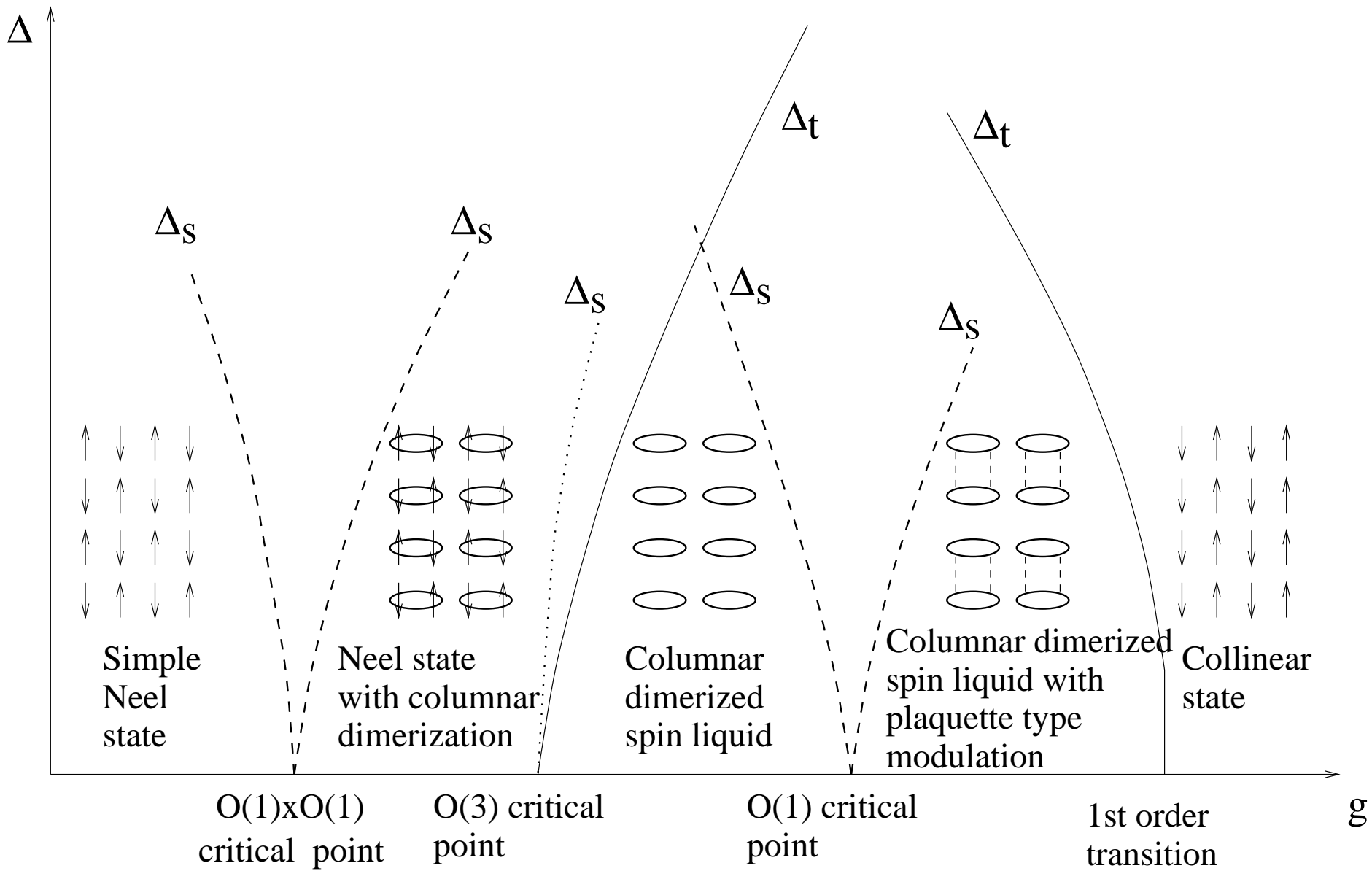
\begin{figure}
\caption
{Schematic phase diagram and the excitation spectra of the $J_1-J_2$ model.
Solid lines show the triplet gap, the dashed lines show the gaps of the {\it relevant} 
singlets, and the dotted line shows the gap of the {\it irrelevant} singlet.}
\label{fig.5}
\end{figure}


\begin{references}
\bibitem{com1} A possibility $g_{c2} < g_{c1}$ has been excluded by
the analysis \cite{us}.
\end{references}
\end{document}